\pdfoutput=1
\documentclass[notitlepage, twocolumn]{revtex4-1}

\usepackage{amssymb}
\usepackage{graphicx}

\begin{document}

\title{Ultrafast Energy Relaxation in Single Light-Harvesting Complexes}

\author{Pavel Mal\'{y}$^{[a,b]}$, J. Michael Gruber$^{[a]}$, Richard J.
Cogdell$^{[c]}$, Tom\'{a}\v{s} Man\v{c}al$^{[b]}$, and Rienk van
Grondelle$^{[a]}$}

\affiliation{$^{[a]}$Department of Physics and Astronomy, Faculty of Sciences,
Vrije Universiteit Amsterdam, De Boelelaan 1081, 1081HV Amsterdam,
The Netherlands, $^{[b]}$Institute of Physics, Charles University
in Prague, Ke Karlovu 5, 12116 Prague, Czech Republic, $^{[c]}$Institute
of Molecular, Cellular and Systems Biology, College of Medical, Veterinary
and Life Sciences, University of Glasgow, Glasgow G128QQ, United Kingdom}
\begin{abstract}
Energy relaxation in light-harvesting complexes has been extensively
studied by various ultrafast spectroscopic techniques, the fastest
processes being in the sub-100 fs range. At the same time much slower
dynamics have been observed in individual complexes by single-molecule
fluorescence spectroscopy (SMS). In this work we employ a pump-probe
type SMS technique to observe the ultrafast energy relaxation in single
light-harvesting complexes LH2 of purple bacteria. After excitation
at 800 nm, the measured relaxation time distribution of multiple complexes
has a peak at 95 fs and is asymmetric, with a tail at slower relaxation
times. When tuning the excitation wavelength, the distribution changes
in both its shape and position. The observed behaviour agrees with
what is to be expected from the LH2 excited states structure. As we
show by a Redfield theory calculation of the relaxation times, the
distribution shape corresponds to the expected effect of Gaussian
disorder of the pigment transition energies. By repeatedly measuring
few individual complexes for minutes, we find that complexes sample
the relaxation time distribution on a timescale of seconds. Furthermore,
by comparing the distribution from three long-lived complexes with
the whole ensemble, we demonstrate that the ensemble can be considered
ergodic. Our findings thus agree with the commonly used notion of
an ensemble of identical LH2 complexes experiencing slow random fluctuations.
\end{abstract}
\maketitle

\section*{Introduction}

Time-resolved studies of primary events in photosynthetic light harvesting
have a decades-long tradition. Usually, the fastest processes observed
correspond to the time resolution of the experimental techniques available
at the time. Recently, the most popular tool to study ultrafast excitation
energy transfer with sub-100 fs resolution is two-dimensional electronic
spectroscopy (2DES). This technique has been used to study various
light-harvesting complexes such as LH2 and LH1 antennas of purple
bacteria\cite{Harel2011,Maiuri2015}, the FMO protein of green sulphur
bacteria\cite{Brixner2005,Engel2007} and the major antenna complex
LHCII of higher plants\cite{Schlau-Cohen2009a}. It was shown that
after an ultrafast excitation of photosynthetic light-harvesting complexes
(LHCs) the electronic excitation evolves in a coherent fashion on
a 100 fs timescale. These observations sparked a still ongoing debate
on the role of quantum coherence in energy transfer in LHCs. 

However powerful the ultrafast techniques have become, they are fundamentally
limited by ensemble averaging. Although the 2DES can in principle
resolve inhomogeneous and homogeneous lineshapes, the observed spectra
and system dynamics are still averaged over the whole ensemble of
complexes. Another feature of nonlinear spectroscopy such as 2DES
is that broadband pulses are used for excitation, which results in
simultaneous excitation of many states. Such pulses inevitably excite
also superpositions of states, which leads to coherent dynamics. This
can provide useful information about the system, especially on the
electronic coupling between the pigments and the interplay of electronic
and nuclear degrees of freedom. On the other hand, it brings with
itself interpretative issues in relation to the relevance of such
coherent dynamics for natural light harvesting under incoherent sunlight\cite{Mancal2010}. 

At about the same time as ultrafast spectroscopy, also optical microscopy
has seen significant advances\cite{Brinks2014}. Nowadays it is routinely
possible to selectively excite and observe individual LHCs. This enables
us to overcome the problem of ensemble averaging and observe distributions
of single-molecule properties. However, for practical reasons only
single-molecule emission spectroscopy has been possible on biological
pigment-protein complexes. Photon counting of the weak luminescence
signal becomes a limiting factor for the time resolution, making it
possible to observe changes only on a timescale of tens of milliseconds
and longer.The standard paradigm is therefore to think about the ultrafast
nonlinear ensemble spectroscopy and single-molecule spectroscopy (SMS)
as complementary methods that access very different timescales. 

In 2005 van Dijk et al. proposed a modification of SMS called single-molecule
pump-probe (SM2P), which employs excitation by two pulses. This technique
visualizes the initial ultrafast excitation relaxation in single molecules\cite{VanDijk2005}.
As they demonstrated on dye monomers\cite{VanDijk2005,VanDijk2005a}
and later on dye dimers and trimers\cite{Hernando2006}, it is possible
to observe relaxation rates in the 100 fs range. In this work we explore
the possibility of applying this technique to light-harvesting complexes. 

The LHC of choice for our measurement is the light-harvesting complex
2 (LH2) of the purple bacterium \textit{Rhodopseudomonas acidophila}.
LH2 consists of two rings of bacteriochlorophylls, which result in
two distinct absorption bands at roughly 800 and 850 nm, respectively.
Both the bands and the rings are referred to as B800 and B850 according
to their central absorption wavelength. The pigments in the ring
responsible for the B800 band are relatively weakly coupled, while
the pigments from the B850 ring exhibit strong electronic coupling.
This strong interaction results in significant excitonic splitting
and formation of delocalized excitonic (vibronic) states\cite{Sundstrom1999}.
Most of the ultrafast studies of energy transfer in LH2 were carried
out in the late eighties and nineties using variants of transient
absorption (TA) and fluorescence  upconversion\cite{Sundstrom1999,Bergstroem1988,Hess1993,Jimenez1996}.
From these and later studies\cite{Wendling2003,Novoderezhkin2003}
it was concluded that while the energy transfer from the B800 to the
B850 ring is relatively slow, 1-2 ps, the relaxation dynamics after
800 nm excitation are more complex, including faster components due
to the overlap of the B800 states with high energetic exciton states
of the B850 ring (B850{*}). These states were found to exhibit ultrafast
transfer dynamics on the timescale of hundreds of fs. Recent results
from 2DES spectroscopy furthermore revealed ultrafast sub-200 fs dynamics\cite{Zigmantas2006,Harel2011,Fidler2014}.
Meanwhile, SMS studies of LH2 at cryogenic and later at ambient temperatures
showed intensity fluctuations and spectral diffusion on a much slower
timescale of seconds\cite{VanOijen2000,Rutkauskas2004,Baier2008,Schlau-Cohen2013}.
By theoretical modeling it was shown that most of the spectroscopic
observations can be explained by dynamic variations in the realization
of the energetic disorder of the pigments\cite{Rutkauskas2004,Rutkauskas2006}.
These findings highlight the dynamic, fluctuating nature of LHCs.
Experimentally, LH2 is a perfect candidate for our proof-of-principle
measurement for several reasons. The presence of lower B850 states
results in fluorescence emission around 870 nm, which is sufficiently
red-shifted with respect to the absorption bands to enable easy excitation
and detection separation. Importantly, LH2 shows a high fluorescence
yield and significant stability in single-molecule conditions, which
is a requirement for our experiment.

\section*{Results}

\subsection*{The measurement}

\begin{figure}
\noindent \includegraphics[scale=0.9]{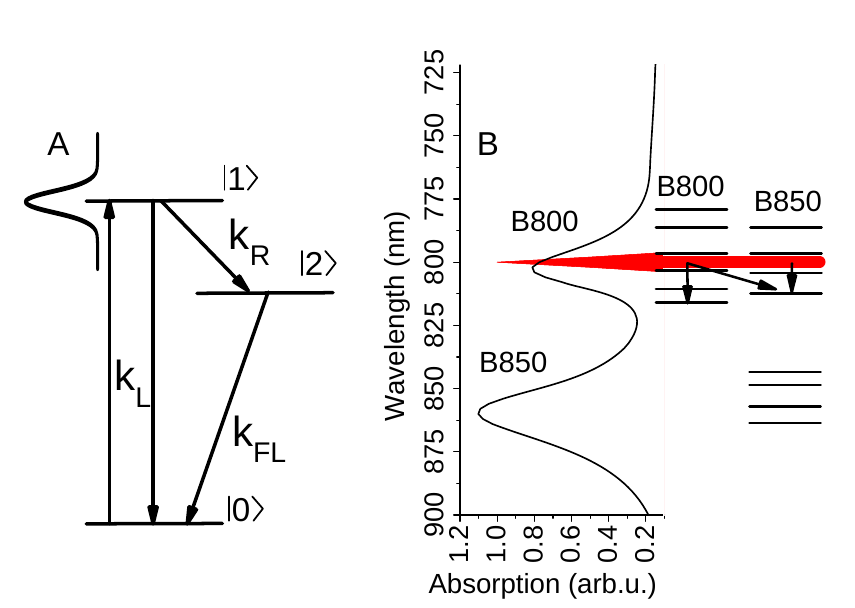}

\caption{(A) The three-level scheme used for the data analysis. $k_{L}$ is
the absorption and stimulated emission rate, $k_{FL}$ is the spontaneous
emission rate and $k_{R}$ is the relaxation rate which is measured.
The Gaussian profile represents the laser pulse, resonant with state
$\left|1\right\rangle $ and off-resonant with state $\left|2\right\rangle $.
(B) Excited states available in LH2, schematically shown together
with a measured absorption spectrum. The red peak represents the excitation
spectrum at 800 nm, the arrows indicate possible relaxation channels.
\label{fig:System-levels}}
\end{figure}

The SM2P principle is based on exciting the system by a near-saturating
laser pulse and giving it a time window to relax to some off-resonant
state before applying a second pulse. By such relaxation the excitation
in the system can be saved from the stimulated emission caused by
the second pulse, and the overall excitation probability therefore
rises with the pulse delay. The detected fluorescence signal is proportional
to this excitation probability and therefore depends on the delay
between the two pulses. The excitation relaxation rate can then be
extracted by scanning the pulse delay time and fitting the resulting
change in fluorescence intensity. The effective three-level scheme
which is used for the SM2P traces analysis is shown in Fig. \ref{fig:System-levels}A.
It consists of a ground state $\left|0\right\rangle $, an excited
state $\left|1\right\rangle $ resonant with the laser excitation
and an off-resonant excited state $\left|2\right\rangle $. This scheme
is universal for the technique and can always be used for analysis.
It then depends on the measured system how the respective levels should
be interpreted. A cartoon of the actual situation in LH2, together
with the measured absorption spectrum, is presented in Fig. \ref{fig:System-levels}B.
The main difference between the isolated molecules studied previously
in Ref. \cite{VanDijk2005} and LHCs is the dense excited states manifold
in the latter case. However, it can be shown by numerical simulations,
that the three-level description still holds as effective. In the
case of a dense manifold, the observed relaxation rate is the effective
rate with which the excitation escapes the region resonant with the
laser. For a more detailed description of the SM2P technique and the
analysis procedure we refer the reader to the supporting information
(SI) and the original works by van Dijk et al.\cite{VanDijk2005,VanDijk2005a}.

\noindent 

\begin{figure}
\includegraphics[scale=0.9]{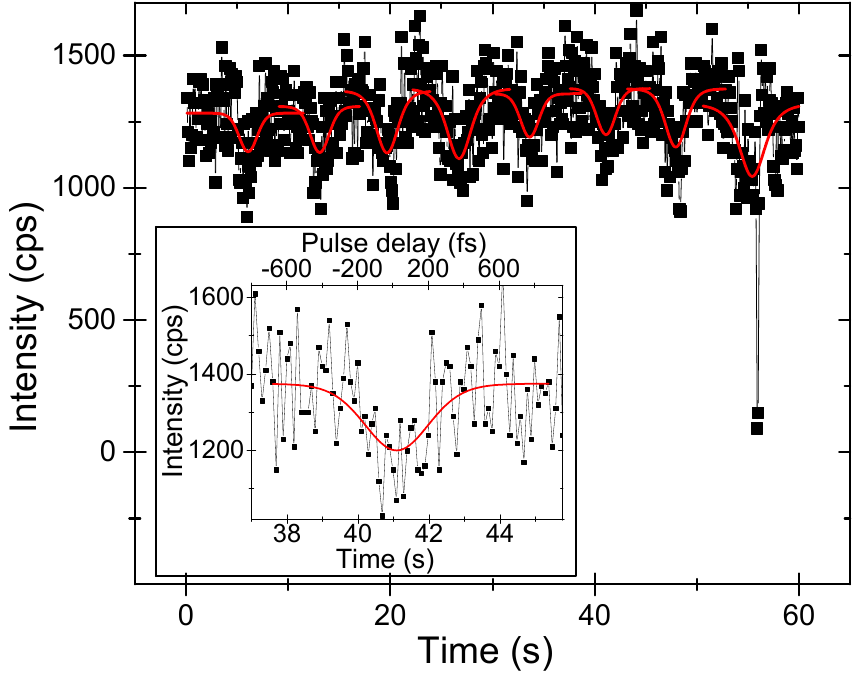}

\caption{First one minute of a measured fluorescence intensity trace of a single
LH2 complex, recorded while continuously scanning the delay between
the two excitation pulses. Red lines: data fitted with the three level
model in Fig. \ref{fig:System-levels}A. At $t=55\mbox{ s}$ the complex
briefly switches to a dark state ('blinking'). Inset: magnification
of one intensity dip with a fitted relaxation time of $\tau_{R}=\left(89\pm25\right)\mbox{ fs}$.
Bottom axis gives the real recording time, while the top axis denotes
the delay between the two pulses.\label{fig:Single-LH2-trace}}
\end{figure}

Using a confocal microscope, individual complexes are excited by the
two-pulse laser sequence. The pulses with a center wavelength around
800 nm are 200 - 250 fs long and about 4 nm wide. Thorough preliminary
calculations, which can be found in the SI, indicated that the above
mentioned laser specifications will work to reveal ultrafast dynamics
in LH2 complexes. The fluorescence of one complex is collected by
the same microscope objective and recorded by an avalanche photodiode.
In this way the fluorescence intensity traces of multiple individual
LH2 complexes are recorded one by one. The emission of one complex
is measured until it photobleaches, while simultaneously continuously
scanning the delay between the two excitation pulses. The first minute
of a typical intensity trace from a stable complex is shown in Fig.
\ref{fig:Single-LH2-trace}. The signal of about 1000 counts per second
is characteristic for the given measurement conditions. The data are
binned into 100 ms bins, which represents a compromise between the
signal-to-noise ratio and the amount of data points available for
fitting. The measured intensity modulation results from the pulse
delay scanning and each intensity dip can be used to determine the
corresponding relaxation time. The inset in Fig. \ref{fig:Single-LH2-trace}
depicts a single intensity dip from the recorded trace, with an extracted
relaxation time of $\tau_{R}=\left(89\pm25\right)\mbox{ fs}$. The
present noise can be explained by Poissonian shot noise. The good
sample stability allowed us to perform multiple pulse delay scanning
cycles and therefore to extract multiple subsequent relaxation times
from one complex. In the given example, the complex switches into
a dark state at $t=55\mbox{ s}$, a process often called 'blinking'.
This behaviour indicates that the observed signal indeed arises from
a single well connected antenna. It should be noted that not all complexes
are such stable emitters. As was observed before (see e.g. \cite{Rutkauskas2004,Schlau-Cohen2013}),
there can be a significant amount of blinking with different degrees
of quenching, which results in switching between different intensity
levels. However, no matter what the mechanism of energy dissipation
causing these fluctuations is, the fluorescence intensity is still
proportional to the excitation probability. Therefore, whenever the
emission is stable for a sufficiently long time to perform one pulse
delay scan, and the emission intensity is high enough to provide a
reasonable signal-to-noise ratio, the relaxation time can be measured.
In this way we can measure several intensity dips for many complexes
and extract the relaxation times by fitting with the three-state model. 

\subsection*{Energy relaxation}

\begin{figure}
\includegraphics[scale=0.9]{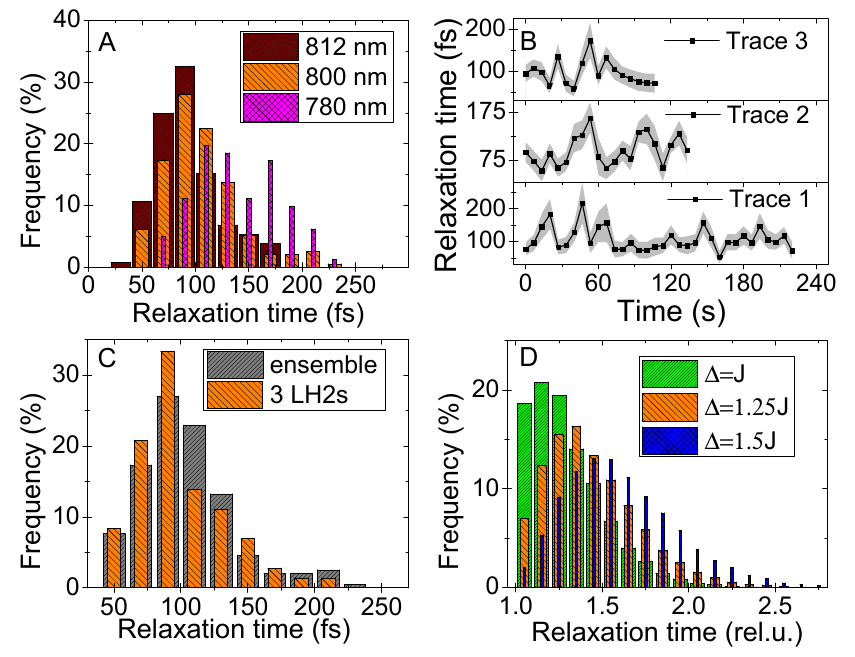}

\centering{}\caption{(A) Relaxation time distribution obtained from many measured complexes
at three different excitation wavelengths (B) Relaxation time trajectories
of three stable complexes, under 800 nm excitation. The shaded regions
indicate the standard error of the fits. (C) The relaxation time distribution
obtained from these 3 complexes, compared to the ensemble distribution
at 800 nm excitation from (A). (D) Modelled distribution of the relaxation
times in a two-state model using Redfield theory. The distribution
shape changes with the ratio of the coupling and energy gap (see text
for description).\label{fig:Results}}
\end{figure}

The distributions of relaxation times obtained for excitation wavelengths
of 812 nm, 800 nm, and 780 nm are shown in Fig. \ref{fig:Results}A.
The average recorded relaxation times are 92 fs at 812 nm, 106 fs
at 800 nm and 139 fs at 780 nm excitation. These measured relaxation
times agree well with the expected ultrafast timescale.

As a result of the already mentioned dense excited states manifold,
there are several differences between the original work on dye monomers
and the LHCs. In the former case of individual or weakly coupled pigments,
the observed ultrafast relaxation is the intramolecular vibrational
relaxation, i.e. the dynamic Stokes' shift. By comparing monomers
and dimers, van Dijk et al. showed that this relaxation slows down
when the excited states are delocalized and thus more weakly coupled
to the environment\cite{VanDijk2005}. In LHCs the situation is different.
First, the pigments are coupled and thus the vibrational and electronic
states become mixed, resulting in a vibronic states manifold. The
energy transfer between these states cannot be strictly separated
into the intra- or inter- pigment relaxation. Second, unlike the dye
molecules, the bacteriochlorophylls present in LH2 have a much smaller
Stokes' shift, typically around 5 nm ($\approx80$ cm$^{-1}$)\cite{DeCaro1994}.
It is thus by itself not enough to escape the 4 nm ($\dot{\approx65\mbox{ cm}^{-1}}$)
wide excitation pulse. And finally, the measured dependence of the
relaxation time on the excitation wavelength is exactly opposite from
what would be expected from a Stokes' shift. In our case we observe
the fastest relaxation in the 'red' region with wavelength longer
than 800 nm, where the strongly-coupled B850{*} states are present.
The relaxation is then slower when exciting in the 'blue' region at
780 nm, where the states of weakly-coupled B800 pigments play a larger
role, see Fig. \ref{fig:Results}A.

Another aspect to consider is that we observe only energy relaxation
and not dynamic localization, because of excitation with circular
polarized light. The latter contributes mainly to absorption depolarization\cite{Novoderezhkin2003}.
The observed relaxation rate then effectively describes how fast the
excitation escapes the resonant laser excitation range. The next dissimilarity
from the case of individual dye molecules is the possible presence
of multiple excitations and the related singlet-singlet annihilation.
However, because the fluorescence lifetime is orders of magnitude
longer than the singlet-singlet annihilation time\cite{VanGrondelle1983,Ma1997},
it is precisely the annihilation which renders the multiply-excited
states invisible. The annihilation, always present at near-saturating
intensities, thus effectively ensures that the three state model with
a single excited state is a good approximation for the observed fluorescence
signal.  Another concern are higher excited states of the pigments,
possibly resulting from multiple excitation of the same pigment. However,
these decay to the lowest excited state much faster than the overall
excited state lifetime\cite{BlankenshipBook}. From the discussion
above we can therefore conclude that we indeed observe energy relaxation
between the singly-excited states within the complex. 

Comparing with the literature, we find that our relaxation times are
somewhat shorter than those found by previous time-resolved measurements.
As mentioned in the introduction, excitation at 800 nm results in
populating states of both the B800 and the B850{*} bands. Our experimental
results indeed indicate that the B850{*} states contribute significantly
to the rather fast observed relaxation rate. The comparably wider
excitation pulses of typically 10-15 nm used in TA measurements fail
to resolve energy relaxation processes within their bandwidth, resulting
in a slower overall relaxation rate. Furthermore, TA and fluorescence
decay kinetics are usually fitted with and resolved into several energy
transfer components, while this study yields an effective 'escape'
rate comprising all available relaxation channels. As a consequence,
the observed relaxation is somewhat faster and the slow components
are not visible in our measurement. Our results therefore agree with
relaxation times of 150-300 fs reported for 800 nm excitation\cite{Hess1993}
and furthermore experimentally validate the faster dynamics determined
by theoretical modeling of the B850{*} band\cite{Novoderezhkin2003}.

\subsection*{Relaxation time fluctuations}

Having discussed the average observed relaxation time, we can focus
on the true single-molecule measurement achievements: the relaxation
time distribution and fluctuations. We have already mentioned the
distributions in Fig. \ref{fig:Results}A. Due to the anaerobic conditions
which increase the sample endurance, we were able to follow several
stable complexes for minutes before they photobleached. The obtained
relaxation time trajectories can be found in Fig. \ref{fig:Results}B. 

\begin{figure}
\includegraphics[scale=0.9]{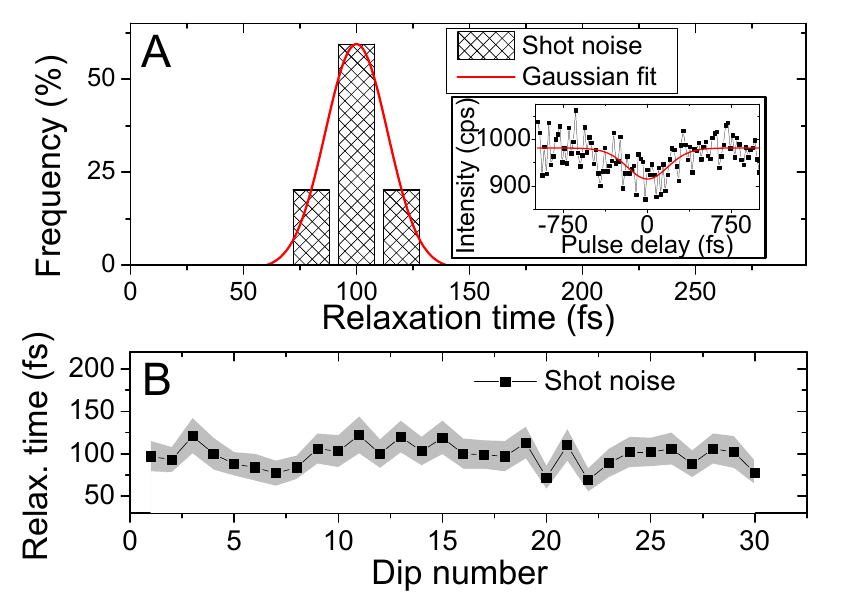}

\caption{Testing the effect of Poissonian shot noise. The simulated parameters
are: a relaxation time of 100 fs, pulses of 200 fs and a signal of
around 1000 cps. (A) The relaxation time distribution obtained from
the simulation, fitted with a Gaussian distribution with a FWHM of
33 fs. Inset: one of the simulated intensity dips, together with the
fitted 3-level model curve. The recovered relaxation time was $\tau_{R}=\left(90\pm17\right)\mbox{ fs}$.
(B) A succession of simulated relaxation times that can be compared
with Fig. \ref{fig:Results}B. The shaded area indicates the standard
error of the fits. \label{fig:poissonTest}}
\end{figure}

Before we start interpreting these results, we need to make sure that
the fluctuations we measure are not just an artifact of the fitting
in the presence of shot noise. To this end we perform numerical simulations
of the SM2P signal including Poissonian shot noise. In the inset
of Fig. \ref{fig:poissonTest}A we present one of the simulated intensity
dips. The distribution of the fitted relaxation times obtained from
such simulated dips is presented in Fig. \ref{fig:poissonTest}A.
The signal binning time and the bin size are the same as used in Fig.
\ref{fig:Results}A to illustrate the difference. The calculated relaxation
times are symmetrically distributed around the expected value of 100
fs and the distribution can be excellently fitted by a Gaussian normal
distribution with a FWHM of 33 fs. This distribution is much narrower
than the experimentally obtained one and also its shape is completely
different. In Fig. \ref{fig:poissonTest}B we show a 'trace' of successively
simulated relaxation times that can be compared to its experimental
counterpart in Fig. \ref{fig:Results}B. The extent of the fluctuations
caused only by the shot noise is significantly smaller. Together with
a clear wavelength dependence of the relaxation time distributions,
these simulations convince us that the observed fluctuations are real
and not only the result of shot noise.

In order to qualitatively understand the possible origin of the asymmetric
shape of the relaxation time distribution, we can consider the following
simple model. Let us describe energy transfer between two excitonic
states, originating from two coupled pigments. Using Redfield theory,
the relaxation rate $k_{rel}$ between the excitonic levels can be
expressed analytically. When we assume, for the sake of simplicity,
that the spectral density of bath modes is approximately flat in the
considered frequency region, the relaxation rate is proportional to
\begin{equation}
k_{rel}=\frac{1}{\tau_{rel}}\propto\frac{1}{1+\left(\frac{\Delta}{2J}\right)^{2}},
\end{equation}

\noindent where $J$ is the coupling constant between the pigments
and $\Delta$ is the energy difference between the coupled states.
The relaxation time is then determined only by the ratio $\frac{\Delta}{2J}$
and the distribution arises from the energetic disorder in $\Delta$.
We assume a Gaussian distributed disorder, as is commonly done in
such simulations, with a FWHM $\Delta_{dis}=J$. This is typical for
simulations of light-harvesting complexes, where all three parameters
are expected to be in the same range, i.e. $\Delta\approx J\approx\Delta_{dis}$.
In Fig. \ref{fig:Results}D the resulting relaxation time distribution
is depicted for different values of the detuning $\Delta$. We can
see that for strong coupling (or small energy gaps) the relaxation
is fastest and the distribution is highly asymmetric. With decreasing
coupling (or increasing energy gap) the distribution maximum shifts
to longer relaxation times and becomes more symmetric. Our experimentally
obtained distributions in Fig. \ref{fig:Results}A seem to follow
this trend: the distribution measured at 812 nm is the most asymmetric
one with the shortest relaxation times, the 800 nm distribution is
the intermediate case and the distribution measured at 780 nm excitation
is more symmetric and shifted to longer relaxation times. This fully
agrees with the discussion above, describing the increasing influence
of the strongly-coupled B850{*} ring states when tuning the excitation
to longer wavelengths. The shape of the distribution can thus be qualitatively
described as originating from a Gaussian energetic disorder of the
transitions energies of the antenna pigments. 

\noindent

Finally, we want to comment on the relaxation time trajectories presented
in Fig. \ref{fig:Results}B. The relaxation time clearly varies on
a timescale of seconds, which is in agreement with slow fluctuations
observed by SMS on LH2 before\cite{VanOijen2000,Rutkauskas2004,Rutkauskas2006}.
It should be mentioned that fluctuations in LHCs are observed on almost
all timescales, from fast sub-picosecond vibrations of the pigments
to slow protein structural changes in the range of seconds. Our experiment
is able to observe the latter type of fluctations, where slow motion
of the protein causes changes in the local pigment environment resulting
predominantly in a shift of their transition energy \cite{Rutkauskas2006,Kruger2010b,Kruger2011b}. 

A question arises whether all complexes are identical and experience
the same fluctuations, or whether the ensemble is heterogeneous. To
investigate this we compare the relaxation time distribution from
three long traces with the one from the whole ensemble of many complexes.
We find, as is shown in Fig. \ref{fig:Results}D, that the distributions
are very similar. This indicates that every complex can likely sample
all the possible relaxation times on a timescale of seconds and that
the ensemble can be considered ergodic. As this argument is not completely
conclusive, further investigation in this direction would certainly
be of interest.

\section*{Conclusions}

We have successfully applied the SM2P technique to LH2 complexes of
purple bacteria. We have demonstrated that it is possible to observe
ultrafast energy relaxation in individual light-harvesting complexes.
As such, our work highlights a new possible way to study photosynthetic
light-harvesting. We have shown how the relaxation time distribution
changes when tuning the excitation wavelength. The observed behaviour
can be explained by varying influence of the B800 and B850{*} states
of the LH2 rings, in agreement with previous ultrafast spectroscopy
studies. By a numerical calculation we were able to qualitatively
explain the shape of the relaxation time distribution as a result
of the energetic disorder of the LH2 pigments. The extent of disorder
corresponds to the values commonly used in bulk spectroscopy modelling.
Our method can be extended to include a detailed excitation wavelength
scan, which would enable us to study energy transfer dynamics of single
LH2 complexes to an extent similar to bulk transient absorption measurements.
Finally, we observed the evolution of the relaxation rate of individual
complexes in time. In accordance with previous SMS studies, we attribute
its fluctuations to slow protein motion, based on the relevant timescale.
Our results thus not only serve as a proof-of-principle measurement
for the SM2P technique on photosynthetic systems, but also present
a fitting piece of evidence to the puzzle of light-harvesting dynamics
in the ever fluctuating antenna complexes.

\subsection*{Materials and methods}

\paragraph*{{\small{}Experimental setup}}

The experimental setup is similar to the one in Ref. \cite{VanDijk2005},
briefly a 76 MHz Ti:Sapphire laser (Mira 900F, Coherent) is used
as a source of 200-250 fs, 4-5 nm spectrally wide pulses centered
at 800 nm. By tuning the laser cavity the wavelength can be tuned
approx. from 750 nm to 850 nm. The repetition rate is decreased to
2 MHz by a pulse-picker (PulseSelect, APE) to increase the survival
time of the complexes and eliminate long-living dark states such as
triplet states. The absence of the triplet states is verified by checking
the signal drops to half when halving the repetition rate. The two
pulses are produced by a home-built Michelson interferometer, the
delay between them is scanned by a delay line (Newport) in one of
the interferometer arms. The pulse length before the microscope is
measured by fringe-resolved autocorrelation\cite{Diels1985} using
the same interferometer and focusing the pulses into a BBO crystal
(Eksma optics). The pulse spectrum is measured by a spectrometer (OceanOptics).
Technical details can be found in the SI. Due to the narrow bandwidth
of the pulses no significant broadening of the pulses in the microscope
can be expected. Guild et al. measured the dispersion of common high
N.A. objective microscopes, and for a microscope very similar to ours
they find GDD of around 4000 fs$^{2}$, including the beam expander
\cite{Guild1997}. Using a formula for Gaussian pulse second-order
dispersion, we obtain that our 200 fs (lower limit) pulses stretch
to 208 fs. This is indeed negligible considering the fluorescence
intensity dip fitting error arising from the signal to noise ratio.
The excitation light is adjusted to a circular polarization by a Berek
compensator (New Focus) to avoid complex orientation dependence. The
complexes are illuminated and detected by a confocal microscope with
a PlanFluor objective (1.3NA, Nikon) as described elsewhere\cite{Rutkauskas2004}.
The detected fluorescence is alternatively dispersed by a grating
on a CCD (Princeton Instruments) to measure the emission spectrum
or the intensity is measured by an avalanche photodiode (Perkin-Elmer).
The fluorescence spectrum is used to check the integrity of the complexes
during the course of the measurement. The excitation intensity is
set to be sufficient to nearly-saturate the complexes. At 800 nm excitation
we used an excitation power of 0.5 pJ/pulse, focused to a diffraction-limited
spot, which is comparable to previous experiments \cite{VanDijk2005a}.
For excitation at different wavelengths the intensity was increased
to compensate for the decreased absorption, see spectrum in \ref{fig:System-levels}B.
The measurement is controlled by a custom-made LabView environment.

\paragraph*{{\small{}Sample preparation}}

The isolated LH2 complexes from \textit{Rhodopseudomonas acidophila}
are diluted to a concentration of $\sim10$ pM in a measuring buffer
(20 mM Tris, pH 8 and 0.03\% (w/v) n-Dodecyl ${\textstyle \beta}$-D-maltoside)
and then immobilized on a PLL (poly-L-Lysine, Sigma) coated cover
glass. The dilution is chosen such as to obtain on average approximately
10 complexes per 100 $\mu$m$^{2}$. To increase the survival time
of complexes the buffer is deoxygenated by the oxygen-scavenging system
PCA/PCD (2.5 mM protocatechuic acid, 25 nM protocatechuate-3,4- dioxygenase,
Sigma)\cite{Swoboda2012}.  The experiments were conducted at room
temperature.

\paragraph*{{\small{}Relaxation time fitting}}

The detailed description of the SM2P technique can be found in the
SI. When applying the three-level system description as in Fig. \ref{fig:System-levels}A,
it can be shown the intensity dip as a function of pulse delay $\tau$
can be described as

{\footnotesize{}
\begin{eqnarray}
I(\tau) & = & I^{\infty}\left\{ 1-\frac{p_{1}}{2-p_{1}}\frac{1}{2}e^{\frac{k^{2}d^{2}}{4}}\left[e^{-k\tau}erfc\left(\frac{1}{2d}\left(d^{2}k-2\tau\right)\right)\right.\right.\nonumber \\
 &  & \left.\left.+e^{k\tau}erfc\left(\frac{1}{2d}\left(d^{2}k+2\tau\right)\right)\right]\right\} ,\label{eq:Fitform}
\end{eqnarray}
}{\footnotesize \par}

\noindent where $k=\frac{1}{\tau_{R}}$ is the relaxation rate, $I^{\infty}$
is the baseline intensity, $p_{1}$ is the probability of excitation
by one pulse ($\frac{1}{2}$ for full saturation) and $d$ is the
effective pulse width, related to the pulse full width at half maximum
(FWHM) as $d=\frac{1}{\sqrt{2ln2}}d_{FWHM}$. We use this formula
to fit the measured dips and extract the relaxation times.\nocite{Loudon2001}

\begin{acknowledgments}

P.M., J.M.G. and R.v.G. were supported by the VU University and by
an Advanced Investigator grant from the European Research Council
(no. 267333, PHOTPROT) to R.v.G.; R.v.G. was also supported by the
Nederlandse Organisatie voor Wetenschappelijk Onderzoek, Council of
Chemical Sciences (NWO-CW) via a TOP-grant (700.58.305), and by the
EU FP7 project PAPETS (GA 323901). R.v.G. gratefully acknowledges
his Academy Professor grant from the Netherlands Royal Academy of
Sciences (KNAW). P.M. and T.M. received financial support from the
Czech Science Foundation (GACR), grant no. 14-25752S. R.J.C. was supported
as part of the Photosynthetic Antenna Research Center (PARC), an Energy
Frontier Research Center funded by the U.S. Department of Energy,
Office of Science, Basic Energy Sciences under Award \#DE-SC0001035. 

\end{acknowledgments}

\setcounter{figure}{0}
\makeatletter  
\renewcommand{\thefigure}{S\@arabic\c@figure} 
\makeatother
\onecolumngrid

\pagebreak{}

\section*{Supplementary Information }

\subsection*{Theoretical considerations}

\begin{figure*}
\begin{centering}
\includegraphics[scale=0.9]{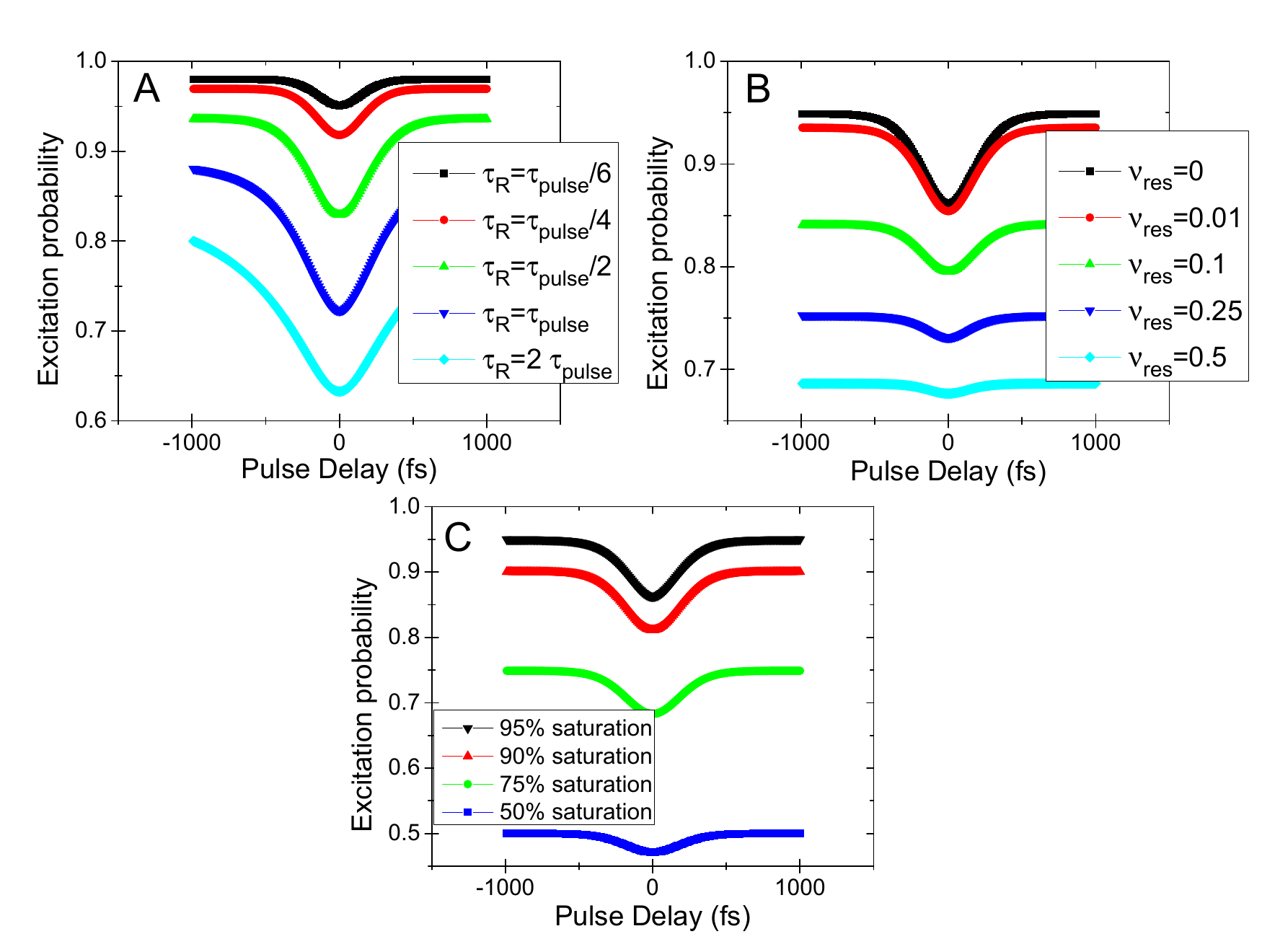}
\par\end{centering}

\caption{Numerical simulations of the intensity dip dependence on (A) the relaxation
time, (B) the off-resonance of the relaxed level and (C) the degree
of saturation. \label{fig: numsims}}
\end{figure*}

The SM2P technique is based on exciting the molecules by two near-saturating
laser pulses, scanning the delay between them, and observing the correlated
change in fluorescence intensity. In order to understand the resulting
signal we study a model three-level system consisting of a ground
state $\left|0\right\rangle $, an excited state $\left|1\right\rangle $
resonant with the laser excitation and an off-resonant excited state
$\left|2\right\rangle $. A schematic illustration is shown in Fig.
1A in the main text. During the interaction with the first pulse the
evolution of the system can be described by the Bloch equations\cite{Loudon2001}.
The external electromagnetic field creates a superposition of the
ground and excited state, i.e. an optical coherence, and the probability
of excitation of state $\left|1\right\rangle $ undergoes Rabi oscillations.
Because of the strong coupling of the electronic degrees of freedom
to the fluctuating environment and fast energy transfer from the excited
state, the optical coherence rapidly dephases, typically on the order
of $\lesssim$50 fs, and the Rabi oscillations become damped. If the
duration of the saturating pulse is longer than the coherence dephasing
time, the excitation probability settles at $\frac{1}{2}$. In that
case a description by semiclassical rate equations  is appropriate.
If the system is excited by the first pulse, it starts to relax to
the off-resonant state $\left|2\right\rangle $. Now, when the second
pulse arrives immediately after the first pulse, the system is still
saturated and the excitation probability remains $\frac{1}{2}$. However,
when the delay between the pulses is long enough, the excitation can
relax to state $\left|2\right\rangle $ and thus escape stimulated
emission. The system can then interact with the second pulse only
if it was not excited by the first pulse. It thus gets a second chance
to be excited and the overall probability of excitation rises to $\frac{1}{2}+\frac{1}{2}\cdot\frac{1}{2}=\frac{3}{4}$.
Scanning the pulse delay and measuring the emission intensity, which
is proportional to the excitation probability, allows us to observe
a dip in fluorescence, see e.g. Fig. \ref{fig: numsims}. The width
of the dip is related to the time it takes for the system to relax
to the off-resonant state. In practice the excitation starts to relax
already during interaction with the exciting pulse, and state $\left|2\right\rangle $
might not be completely off-resonant. The dynamics of the system can
then be described by coupled equations for the state populations

\begin{eqnarray}
\frac{\partial P_{1}(t,\tau)}{\partial t} & = & k_{L}\left(P_{0}-P_{1}\right)I_{L}(t,\tau)-k_{R}P_{1},\nonumber \\
\frac{\partial P_{0}(t,\tau)}{\partial t} & = & -k_{L}\left(\left(1+\nu_{res}\right)P_{0}-P_{1}-\nu_{res}P_{2}\right)I_{L}(t,\tau),\nonumber \\
\frac{\partial P_{2}(t,\tau)}{\partial t} & = & \nu_{res}k_{L}\left(P_{0}-P_{2}\right)I_{L}(t,\tau)+k_{R}P_{1}.\label{eq:rateeqs}
\end{eqnarray}

\noindent Here $k_{L}$ is the absorption/stimulated emission rate,
$k_{R}$ is the relaxation rate, $I_{L}(t,\tau)$ is the intensity
of the two laser pulses delayed by $\tau$, and $\nu_{res}$ is the
resonance of state $\left|2\right\rangle $ (relative to state $\left|1\right\rangle $,
$\nu_{res}=0$ when state $\left|2\right\rangle $ is completely off-resonant).
Because the spontaneous emission rate is typically orders of magnitude
slower than energy relaxation, the observed fluorescence is proportional
to the population of the relaxed state long after interaction with
the exciting pulses, $I_{FL}(\tau)\mbox{\ensuremath{\propto}}P_{2}(\infty,\tau)$. 

Concerning LHCs, several aspects have to be taken into account. Their
states, originating from more than one pigment, form a dense excitonic
(vibronic) manifold and the energy transfer is ultrafast. It is thus
important to verify how sensitive the method is to the actual degree
of saturation, how much off-resonant the state $\left|2\right\rangle $
has to be, and how the fastest observable relaxation rate depends
on the duration of the pulse. To address these questions we performed
numerical simulations varying the respective parameters, see Fig.
\ref{fig: numsims}. It turns out that the exact saturation level
is not critical (Fig. \ref{fig: numsims}C), in agreement with Ref.
\cite{VanDijk2005a}. The off-resonance is, however, very important
(Fig. \ref{fig: numsims}B). For example, already 75\% off-resonance,
i.e. $\nu_{res}=0.25$, decreases the intensity dip considerably and,
in the presence of Poissonian noise (see below), renders it invisible.
Finally, the fastest observable relaxation time is about 5 times faster
than the pulse duration (Fig. \ref{fig: numsims}A). These findings
are very encouraging for LHCs. The expected relaxation times are around
100 fs, which can be well resolved by 200 fs laser pulses. These pulses
can then be only about 4 nm wide, which, together with a high off-resonance
demand, ensures high excitation selectivity. The weak dependence
on the degree of saturation furthermore allows to quantitatively measure
the relaxation out of the selected narrow excitonic region.

The findings presented above allow us to use an effective description
\textit{via} Eqs. (\ref{eq:rateeqs}) with $\nu_{res}=0$. Level $\left|1\right\rangle $
then represents the states resonant with the excitation and level
$\left|2\right\rangle $ the off-resonant, relaxed states. The effective
description by Eqs. (\ref{eq:rateeqs}) is therefore still valid even
for LHCs with dense excited state manifolds. 

For Gaussian pulses the solution of Eqs. (\ref{eq:rateeqs}) can be
found analytically by convoluting the result for $\delta$-pulse excitation
with both pulse envelopes. This approach, the same as in the original
work of van Dijk et al.\cite{VanDijk2005a}, assumes the pulses are
mixed incoherently. This is true for our setup where the vibrations
of the rapidly-moving delay line safely destroy all phase coherence.
We also tried measuring with a vibrating mirror and obtained comparable
results. The measured fluorescence intensity as a function of the
pulse delay $\tau$ can then be expressed as

\begin{equation}
I(\tau)=I^{\infty}\left\{ 1-\frac{p_{1}}{2-p_{1}}\frac{1}{2}e^{\frac{k^{2}d^{2}}{4}}\left[e^{-k\tau}erfc\left(\frac{1}{2d}\left(d^{2}k-2\tau\right)\right)+e^{k\tau}erfc\left(\frac{1}{2d}\left(d^{2}k+2\tau\right)\right)\right]\right\} \label{Fitform-1}
\end{equation}

\noindent where $I^{\infty}$ is the baseline intensity, $p_{1}$
is the probability of excitation by one pulse ($\frac{1}{2}$ for
full saturation), $k$ is the relaxation rate and $d$ is the effective
pulse width, related to the pulse full width at half maximum (FWHM)
as $d=\frac{1}{\sqrt{2ln2}}d_{FWHM}$. Formula (\ref{Fitform-1})
is used for fitting the data and extraction of the relaxation time
$\tau_{R}=\frac{1}{k}$.

\subsection*{Pulse characterization}

To characterize the pulses we use a modified version of fringe-resolved
autocorrelation (FRAC)\cite{Diels1985}. In this interferometric method
we use the same Michelson interferometer as we use for the pulse delay
scanning. For the pulse characterization the pulses are colinearly
focused in a BBO crystal to generate a second harmonic signal (SH).
The fundamental frequency is then filtered out and the SH intensity
is detected by a slow detector. 

Denoting the laser field $E(t,\tau)=E_{0}(t)e^{-i\omega t}+E_{0}(t+\tau)e^{-i\omega(t+\tau)}+c.c.$,
the second harmonic is proportional to $E^{2}(t,\tau)$, and the measured
SH intensity is proportional to

\begin{eqnarray}
I_{SH}(\tau) & = & \int dtI_{0}^{2}(t)+I_{0}^{2}(t+\tau)+2Re\{E_{0}^{2}(t)E_{0}^{2}(t+\tau)\}cos(2\omega\tau)\nonumber \\
 &  & +4\left(I_{0}(t)+I_{0}(t+\tau)\right)Re\{E_{0}(t)E_{0}(t+\tau)cos(\omega\tau)+4I(t)I(t+\tau).\label{eq:ISH}
\end{eqnarray}

\noindent This signal contains a baseline contribution from both pulses,
the SH interferogram, envelope-modulated fundamental interferogram
and the intensity autocorrelation which we want to measure. Using
this FRAC signal the pulses can be characterized, however, for pulses
as long as ours, the recording requires an impractically long time.
By attaching a vibrator to one of the interferometer mirrors, the
interference-sensitive part of the SH is averaged out on the detector,
which enables us to measure the intensity autocorrelation on a constant
background much faster. We checked that this method works by comparing
the obtained pulse width with the one from the full recorded interferogram
for several pulse lengths.

\begin{figure*}
\begin{centering}
\includegraphics[scale=0.9]{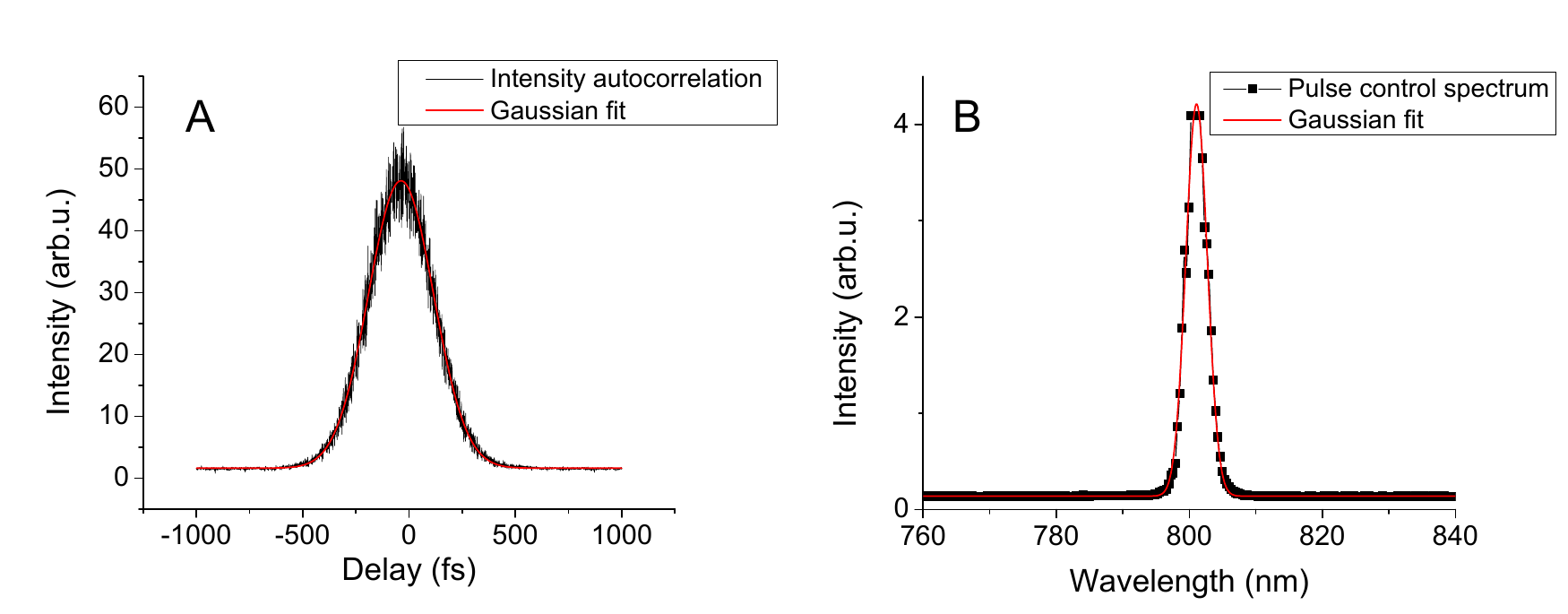}
\par\end{centering}

\caption{Characterization of the pulses used for the 800 nm excitation measurement.
(A) Intensity autocorrelation from using FRAC with vibrating mirror.
Gaussian fit yields $FWHM=360\mbox{ fs}$, using deconvolution factor
$\sqrt{2}$ then gives 255 fs long pulses. (B) Measured control spectrum
(its parameters are not used for the relaxation time extraction).
Gaussian fit gives $FWHM=3.8\mbox{ nm}$ spectral width.\label{fig:Characterization-of-the}}
\end{figure*}

The characterization of the pulses used for the 800 nm excitation
measurement, together with their measured spectrum for control, can
be found in Fig. \ref{fig:Characterization-of-the}.

\noindent 

\noindent 

\end{document}